\documentclass[%
 aip,
 amsmath,amssymb,
 reprint,%
]{revtex4-1}

\usepackage{graphicx, dblfloatfix}
\usepackage{dcolumn}
\usepackage{bm}
\usepackage[utf8]{inputenc}
\usepackage[T1]{fontenc}
\usepackage{mathptmx}
\usepackage{etoolbox}
\usepackage{amssymb}
\usepackage{xcolor}
\usepackage{amsmath}
\usepackage{mathtools}
\usepackage{bm}

\makeatletter
\def\@email#1#2{
 \endgroup
 \patchcmd{\titleblock@produce}
  {\frontmatter@RRAPformat}
  {\frontmatter@RRAPformat{\produce@RRAP{*#1\href{mailto:#2}{#2}}}\frontmatter@RRAPformat}
  {}{}
}
\makeatother
\begin{document}

\preprint{AIP/123-QED}

\title[Dynamical and statistical properties of estimated high-dimensional ODE models: The case of the Lorenz '05 type II model]{Dynamical and statistical properties of estimated high-dimensional ODE models: The case of the Lorenz '05 type II model}

\author{Aljaž Pavšek}
\email{aljaz.pavsek@ijs.si}
\affiliation{Jozef Stefan Institute, Jamova cesta 39, 1000 Ljubljana, Slovenia}
\author{Martin Horvat}
\affiliation{Faculty of Mathematics and Physics, University of Ljubljana, Jadranska ulica 19, 1000 Ljubljana, Slovenia}
\author{Ju\v s Kocijan}
\affiliation{Center for Information Technologies and Applied Mathematics, University of Nova Gorica, Vipavska cesta 13, 5000 Nova Gorica, Slovenia}

\date{\today}

\begin{abstract}
The performance of estimated models is often evaluated in terms of their predictive capability. In this study, we investigate another important aspect of estimated model evaluation: the disparity between the statistical and dynamical properties of estimated models and their source system. Specifically, we focus on estimated models obtained via the regression method, sparse identification of nonlinear dynamics (SINDy), one of the promising algorithms for determining equations of motion from time series of dynamical systems. We chose our data source dynamical system to be a higher-dimensional instance of the Lorenz 2005 type II model, an important meteorological toy model. We examine how the dynamical and statistical properties of the estimated models are affected by the standard deviation of white Gaussian noise added to the numerical data on which the estimated models were fitted. Our results show that the dynamical properties of the estimated models match those of the source system reasonably well within a range of data-added noise levels, where the estimated models do not generate divergent (unbounded) trajectories. Additionally, we find that the dynamics of the estimated models become increasingly less chaotic as the data-added noise level increases. We also perform a variance analysis of the (SINDy) estimated model's free parameters, revealing strong correlations between parameters belonging to the same component of the estimated model's ordinary differential equation.
\end{abstract}

\maketitle

\begin{quotation}
The time evolution of a physical system is understood as a dynamical system. Extracting the parameters of dynamical systems from measurements is a crucial part of everyday practice in many areas of science and engineering. When we create a model that imitates a real-world system from measurements via a numerical procedure, we say that we have determined an estimated model of the
observed system. Typically, estimated models are evaluated based on their ability to make accurate predictions that closely match real data over a given simulation time. However, in this study, we focused on another aspect of estimated model evaluation that is crucial for understanding and appropriate use of estimated models: the differences in dynamical and statistical properties between the estimated models and their source systems, which is the system that generated the numerical data used to build the estimated models. Specifically, we investigated how the noise level in the numerical data that is used for estimated model learning affects the dynamical and statistical properties of estimated models. To conduct our analysis, we used the Lorenz 2005 type II model\cite{L05} (L05 II), a popular toy model in meteorology. Estimated models were obtained using a recently popular SINDy\cite{SINDy} algorithm, which is a sparse regression method that can deduce equations of motion from a given system's time series data. The selection of the L05 II source system was based on its inherent characteristics that make it well-suited for the examination of both long-term and short-term dynamics, as well as its capacity to effectively test the SINDy algorithm in diverse scenarios, such as dimensionality and chaoticity. Our findings suggest that the dynamical properties of the estimated models match those of the source system well within a certain range of data-added noise amplitudes, where the estimated models do not generate unbounded trajectories. Additionally, we observed that as the amplitude of the data-added noise increased, the estimated models exhibited less chaotic dynamics and notably break certain symmetries present in the source system. We also conducted an analysis of the estimated model's free parameters, which revealed strong correlations between parameters belonging to the same component of the estimated model's equation of motion. Overall, this research provides valuable insights into the evaluation of estimated models in capturing the behavior of complex systems. Furthermore, it provides systematic analysis of the widely recognized SINDy estimator, and it represents a contribution to the study of L05 II.
\end{quotation}

\section{INTRODUCTION}\label{ch intro}

In this paper, we want to present the study of the change in dynamical and statistical properties between the dynamical system that is the source of numerical data and the data-driven mathematical model, i.e., an estimated model that is meant to represent the former.

One of the important reasons to obtain an estimated model is its use for the prediction of points along the trajectory of the studied dynamical system. This is a very common problem in many areas of physics and engineering and is generally very challenging, particularly when the system under consideration has nonlinear dynamics and our analytical knowledge of the dynamics of the system is incomplete.

There are various approaches to the prediction of dynamics, such as modal decomposition methods \cite{DMD}, symbolic regression \cite{symbolic-reg}, and machine-learning techniques \cite{machine-lorenz}, each with their own strengths and weaknesses. The applicability of these methods may be restricted by issues such as numerical stability, memory storage, and time efficiency. Furthermore, one significant challenge with some of these methods is the analytical interpretation of the dynamics of the mathematical model, extracted from the numerical data.

The paper's primary focus is to present and analyze an approach that has recently gained popularity in the problem of discovering dynamical systems from numerical data in the form of sparse regression\cite{lai2021finding}, which is also referred to as the compressive-sensing approach, as introduced in the work by Wang et al.\cite{wang2011predicting}. A key motivation for utilizing sparse regression is the realization that physical dynamical systems often possess a parsimonious structure, that is, they can be described by a small number of parameters. Building on this insight, a robust algorithm was developed, referred to as sparse identification of nonlinear dynamics (SINDy)\cite{SINDy}, which can extract sparse dynamical equations from numerical data. Although the SINDy estimator has gained increased attention in recent years following the publication of the influential paper\cite{SINDy} in 2016, it is essential to note that a fundamentally similar approach, known as zeroing-and-refitting, had already been introduced two decades earlier \cite{kadtke1993global}. The effectiveness of SINDy has been demonstrated by reconstructing the system of ordinary differential equations (ODEs) from data collected from various non-linear fluid mechanics models. As an illustration, the method was applied on the model of liquid wave formation behind a cylinder \cite{cylinder-wake} and the well-known chaotic Lorenz (1963) system \cite{lorenz63}, as well as other examples, even in the presence of added noise in the collected data, as presented in the featured paper \cite{SINDy} and other studies.

The SINDy estimator offers several notable advantages in uncovering the underlying dynamics of physical systems from numerical data. One of the key benefits of this method is its low computational time complexity and fast convergence, which allows for efficient processing of large data sets. Additionally, the estimator's robustness to noise in the data ensures that the results obtained are reliable, even in cases when the data are not perfect. Another of the most significant benefits of the SINDy estimator is the interpretability of the results. The system of ODEs obtained via this method provides a clear and transparent representation of the underlying dynamics of the system, in contrast to other popular black-box methods, which may not offer this level of insight. This feature of the SINDy estimator allows for a deeper understanding of the system's behavior, which can be crucial for various applications in physics and engineering.

The system of ODEs identified through the SINDy estimator generates an estimated model, which closely approximates the original or source system and can serve as a viable replacement. The primary objective when creating these estimated models is often to maximize their predictive capability, i.e., the accuracy of their predictions over a given simulation time, given an initial condition. However, this should not be the sole evaluation criterion for an estimated model. In situations where the original dynamical system exhibits strong sensitivity to initial conditions, estimated models might struggle with predictive capabilities. Therefore, it is essential to consider the dynamical and statistical properties of the original system when constructing estimated models. This approach ensures that the estimated models not only provide accurate predictions, but also effectively capture the underlying dynamics and statistical characteristics of the original system.

The contribution of this paper will, therefore, be an investigation of the dynamical and statistical properties of the estimated high-dimensional ODE models. In particular, the focus will be on how the dynamical properties of the estimated models obtained with the SINDy algorithm are affected by the measurement noise\cite{schreiber1996observing}, specifically, by the standard deviation of the white Gaussian noise added to the numerical data on which the estimated models were fitted. We decided to study the dependence of the dynamical properties of the estimated SINDy models on the example of the source system being the Lorenz 2005 type II model \cite{L05} (to which we will refer to as L05 II), which is an important toy model in the field of meteorology. This model has some convenient properties for our study, such as rich dynamics, certain symmetries, arbitrary dimensionality of the system and the restriction of the trajectories to a finite volume of state-space. The construction of estimated models can also be driven by objectives other than making predictions, such as indirectly estimating specific properties of the investigated dynamical system\cite{kadtke1993global}. In these scenarios, the present paper can serve as a reference regarding the extent to which the considered methodology for building estimated models, i.e., the SINDy algorithm, can efficiently extract the dynamical properties of the inspected (high-dimensional) dynamical system.

In real-world situations involving modeling higher-dimensional systems, one frequently encounters the problem of observability, which refers to the issue of measuring only a subset of system variables that are necessary for a complete description of the state in the system's state-space. Addressing this issue often involves state-space reconstruction techniques, such as time-delay embedding and principal component analysis (PCA)\cite{gibson1992analytic}, (delay) differential embedding\cite{lainscsek2015delay}, and more recent approaches using autoencoders\cite{jiang2017state}. Tackling such problems naturally involves numerous challenges \cite{letellier1998non}, with the central one being the determination of the minimal attractor embedding dimension. While there are some important analytical results with limited applicability in this regard\cite{takens2006detecting}, numerical methods are typically employed to partially overcome these limitations\cite{packard1980geometry, kennel1992determining}. Additionally, a crucial concern is identifying a coordinate basis in which the estimated model assumes a sparse representation. While this paper acknowledges the complexities of tackling these issues, it proceeds under the assumption that there are no unmeasured system variables and that a coordinate basis is suitable for performing sparse regression.

To provide a historical context and acknowledge potentially relevant works, it is important to note that the field of nonlinear system identification (with a focus on building sparse global ODE models) has a history spanning over three decades, beginning with the pioneering work by Crutchfield and McNamara\cite{crutchfield1987equations}. A comprehensive overview of this field can be found in the seminal article by Aguirre and Letellier\cite{aguirre2009modeling}. As highlighted by the authors of this work, the (global) modeling of nonlinear dynamical systems has its roots in both engineering and mathematical physics, where the former takes a more practical approach and the latter focuses on autonomous and chaotic systems. Considering this interdisciplinary foundation, we acknowledge that the terminology used in this paper may not align perfectly with the expectations of readers from either end of the spectrum. Nevertheless, we strive to maintain a balanced perspective that bridges the gap between these two fields and accommodates the interests of a broad audience. 

This paper is roughly divided into three parts. In the first part (Sec. \ref{ch methods}), we will discuss the theory needed to address the problem at hand, i.e., we will briefly present the SINDy algorithm, L05 II and its relevant features, and the employed methods of dynamical and statistical analysis. In the second part (Sec. \ref{ch workflow}) we will present the proposed workflow that was carried out to arrive at the results. Last, we will present and discuss the findings of this study in Sec. \ref{ch results}.

\section{METHODS}\label{ch methods}

As we implied in Sec. \ref{ch intro}, the objective of the present research was to evaluate the performance of the SINDy algorithm being subject to noisy data, in terms of the dynamical properties of the sought-after estimated SINDy models. For the reasons that will become clear in the present section, we chose the source dynamical system to be Lorenz 2005 model of type II (L05 II) which has rich dynamics and possesses intriguing properties.

We emphasize that the central focus of this study is on the SINDy algorithm, i.e., L05 II is a carefully picked toy model whose purpose is to serve as a source dynamical system for the testing of the SINDy algorithm. Thus, the results are model-specific and should not be recklessly generalized to the broad field of regression problems where the SINDy algorithm could be applied. Nevertheless, this paper should serve as a caveat to such practices.

The purpose of the present section is to transparently lay down the theoretic footing that is essential to understanding the results and their implications, reported in Sec. \ref{ch results}. Following the main SINDy article \cite{SINDy}, we will first define the sparse regression problem \cite{SINDy2} and briefly present a variant of the basic SINDy algorithm. Second, we will introduce the L05 II system focusing on its properties that play a relevant role in the present study. Last, we will review the chosen methods of dynamical and statistical analysis that were selected as suitable characteristics on the basis of which the source system and the estimated models can be compared.

\subsection{Sparse identification of nonlinear dynamics (SINDy)}\label{ch SINDy}

Sparse identification of nonlinear dynamics (SINDy) \cite{SINDy} is a method developed for the purpose of determining dynamical equations from noisy numerical data collected from numerical simulations or real-world physical experiments. The method assumes that, as in many physically relevant scenarios, the observed system's dynamics can be expressed in the form of a $N$-dimensional ODE (system of ODEs)
\begin{equation}\label{ODE}
\dot{\mathbf{x}}(t)=\mathbf{f}(\mathbf{x}(t)),
\end{equation}

whose right-hand side (RHS), i.e., the vector function $\mathbf{f}$ is in every component made up of only a handful of non-zero terms. That is, if we chose a library $\bm{\Theta}$ of all function terms that could (with respect to some relevant prior knowledge on the problem, such as dimensionality and a coordinate basis) govern the dynamics of our system, we expect the solution $\mathbf{f}$ to be sparse in the space of all such possible functions. 

Suppose that, by being provided with numerical data on the state vector $\mathbf{x}(t)=\begin{bmatrix}x_{0}(t) & x_{1}(t) & \cdots\ & x_{N-1}(t)\end{bmatrix}^{T} \in \mathbb{R}^{N}$ of the system, measured at consecutive time instances $t_i$, $i\in0,\cdots, m-1$, we first construct the matrix of the system states 
\begin{equation}\label{matrix of the system states}
\mathbf{X}=\begin{bmatrix}
\mathbf{x}_0 & \mathbf{x}_1 & \cdots & \mathbf{x}_{m-1}
\end{bmatrix}^T\in\mathbb{R}^{m\times N}.     
\end{equation}

\noindent
The corresponding derivatives matrix, represented as $\dot{\mathbf{X}}=\begin{bmatrix}
\dot{\mathbf{x}}_0 & \dot{\mathbf{x}}_1 & \cdots & \dot{\mathbf{x}}_{m-1}
\end{bmatrix}^T$ can be computed using a suitable technique, the choice of which can directly impact the final models\cite{letellier2009frequently}. In our study, we opted for the Savitzky–Golay method\cite{SavitzkyGolay}, utilizing third-order polynomials for the calculations, which is suitable for working with noisy data. Building the library of function candidates \cite{SINDy} $\bm{\Theta}(\mathbf{X})=\begin{bmatrix}\boldsymbol{\theta}_0(\mathbf{X}) & \cdots & \boldsymbol{\theta}_{P-1}(\mathbf{X})\end{bmatrix}\in\mathbb{R}^{m\times P}$, evaluated on the data $\mathbf{X}$, we seek the solution of equation
\begin{equation}\label{problem eq}
\dot{\mathbf{X}}=\bm{\Theta}(\mathbf{X})\bm{\Xi},
\end{equation}
i.e., we are searching for the matrix of coefficient vectors $\bm{\Xi}=\begin{bmatrix}\boldsymbol{\xi}_0 & \boldsymbol{\xi}_1 & \cdots & \boldsymbol{\xi}_{N-1}\end{bmatrix}$
$\in\mathbb{R}^{P\times N}$ that minimizes the above expression. The aforementioned requirement of $\mathbf{f}$ being sparse in the space of all candidate functions translates into condition on $\bm{\Xi}$ being sparse. Trying to find $\bm{\Xi}$ that best solves the Eq. \eqref{problem eq} numerically translates into minimizing the expression
\begin{equation}\label{minimization}
||\dot{\mathbf{X}}-\bm{\Theta}(\mathbf{X})\bm{\Xi}||_2^2 + \alpha R({\bm{\Xi}})
\end{equation}
over $\bm{\Xi}$, where $||\cdot||_2^2$ denotes the square of the Frobenius norm of a matrix, usually referred to as the $\ell_2$ norm\footnote{As denoted in \cite{SINDy}.}, and $R$ is the chosen regularizer function\cite{horvatsirca}. The strength of regularization is controlled by the scalar parameter $\alpha$. 

A popular method to enforce the sparsity of the solution
$\bm{\Xi}$ is to choose $R$ to be the $\ell_1$ norm, resulting in LASSO\cite{tibshirani1996regression}.
However, LASSO can be computationally inefficient when dealing with exceptionally large data sets\cite{SINDy}. Additionally, prior research has indicated that LASSO may yield models that are not genuinely sparse, as a considerable number of terms in the coefficient matrix $\bm{\Xi}$ can be small yet non-zero\cite{ConsSR3}. Owing to these concerns, the authors of SINDy\cite{SINDy} proposed a so-called sequential thresholded least squares (STLSQ) algorithm, which arrives at the solution in an iterative fashion. At each step, firs, the regularized least squares solution to \eqref{minimization} is computed, where $R=\ell_2$ is chosen.\footnote{The option of $\ell_2$  regularization, i.e., ridge regularization is only presented in the official Python-based SINDy library \cite{pysindy} and not in the initial SINDy paper \cite{SINDy}.}
Then, all the coefficients of $\bm{\Xi}$ that are smaller than some predefined value $\lambda$, are zeroed out. The procedure is iterated until only a handful of terms in $\bm{\Xi}$ are different from zero or until convergence. 

The algorithm is easy on computer memory, turns out to be robust to noisy data and is also fast convergent, usually arriving at the solution only after a couple of iterations \cite{SINDy}. The noise level-dependent performance of the STLSQ algorithm was studied only on generic, low-dimensional systems, focusing mainly on the forecasting abilities and the attractor shape of the resulting estimated dynamical systems. However, it remained unclear how exactly do properties of the estimated models obtained via STLSQ vary with respect to the changing noise level in the data, how the algorithm performs over a range of different, especially higher state-space dimensions, dynamic regimes, and sparsity parameters\cite{SINDy2} of the source system's ODE in the feature space (assuming source system's ODE is known analytically). We suggest that the L05 II case study could be suitable to help to answer some of the above questions.

\subsection{Lorenz 2005 type II model}\label{ch L05}

Lorenz 2005 type II model (L05 II) is a meteorological toy model that represents one-dimensional transport of a scalar quantity on a closed chain, i.e., on a finite set of discrete points with periodic boundary conditions.\cite{L05} It is represented in a form of an ODE (a system of ODEs) whose structure is determined by three model-specific parameters; $N$ is the dimension of the system, $F$ is a forcing constant that impacts the dynamical regime of the system, and the parameter $K$ can be used to control the number of bilinear terms on the RHS of the model's ODE. The equation that defines the time evolution of the state vector $\mathbf{x}$ in its $n$th component is given by
\begin{equation}\label{model2}
\frac{d x_{n}}{d t}=f_n(x_0,\dots, x_{N-1})=[x, x]_{K, n}-x_{n}+F,
\end{equation}
with
\begin{equation}\label{bra}
[x, x]_{K, n}=\frac{1}{K^{2}}\sideset{}{'}\sum_{i=-J}^{J}\sideset{}{'}\sum_{j=-J}^{J}(x_{n-K+j-i} x_{n+K+j}-x_{n-2 K-i} x_{n-K-j}),
\end{equation}
where $n=0,\cdots,N-1$. In the case of $K$ being an even number, $J=K/2$ and $\sum'$ denotes a modified summation where the first and last terms are divided by two. For odd $K$ we perform the usual summation, where $J=(K-1)/2$. While the structure of Eq. \eqref{model2} might appear complex, its components possess a clear physical interpretation. With $x_n$ representing the value of a scalar variable at $n$th node on the chain, if taken to be positive, the constant term represents the forcing, the (negative) linear term indicates the damping of this variable, and the bilinear terms function as convective terms.

\begin{figure}[!h]
\centering
\includegraphics[width=0.235\textwidth]{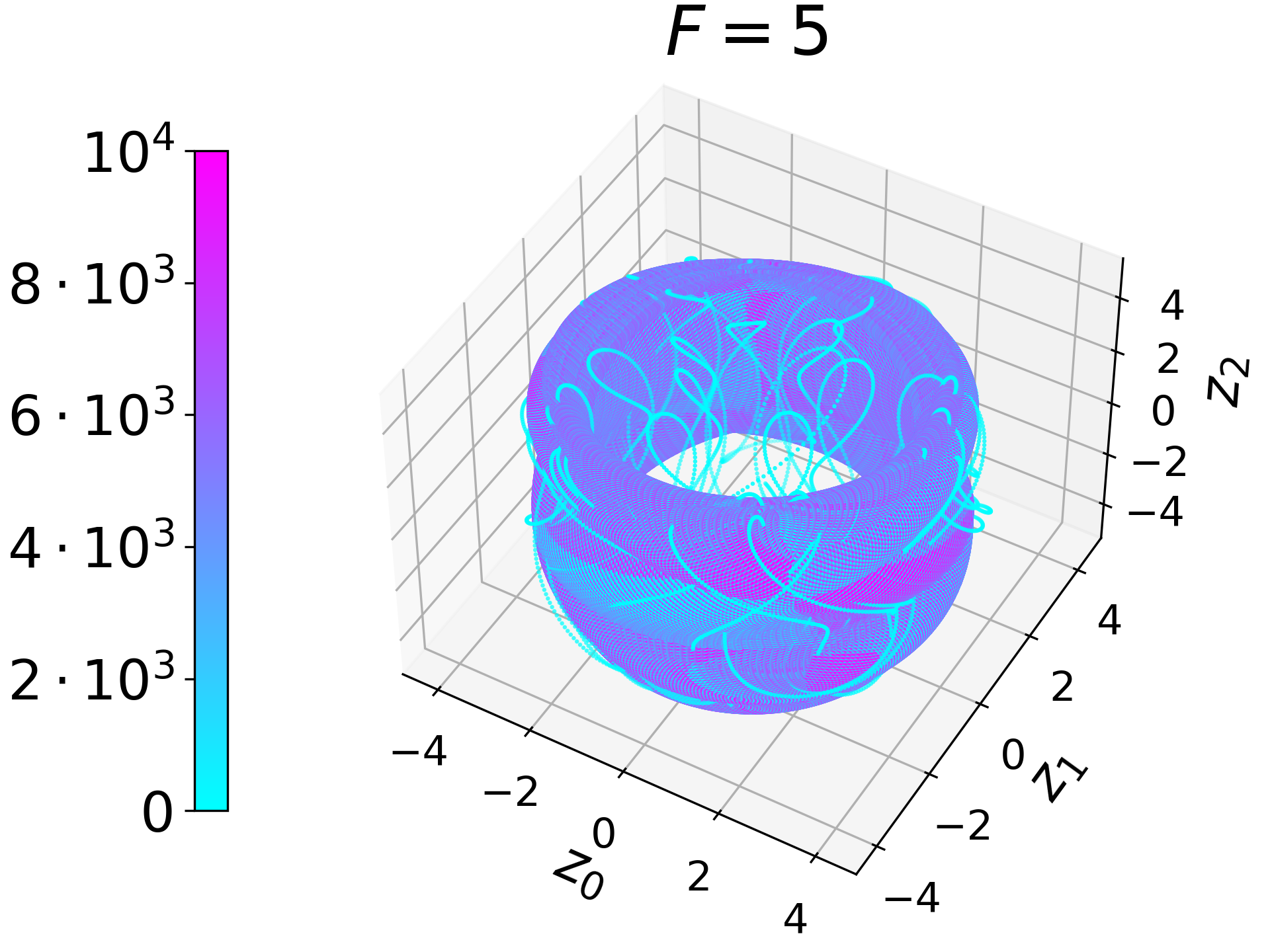}
\includegraphics[width=0.235\textwidth]{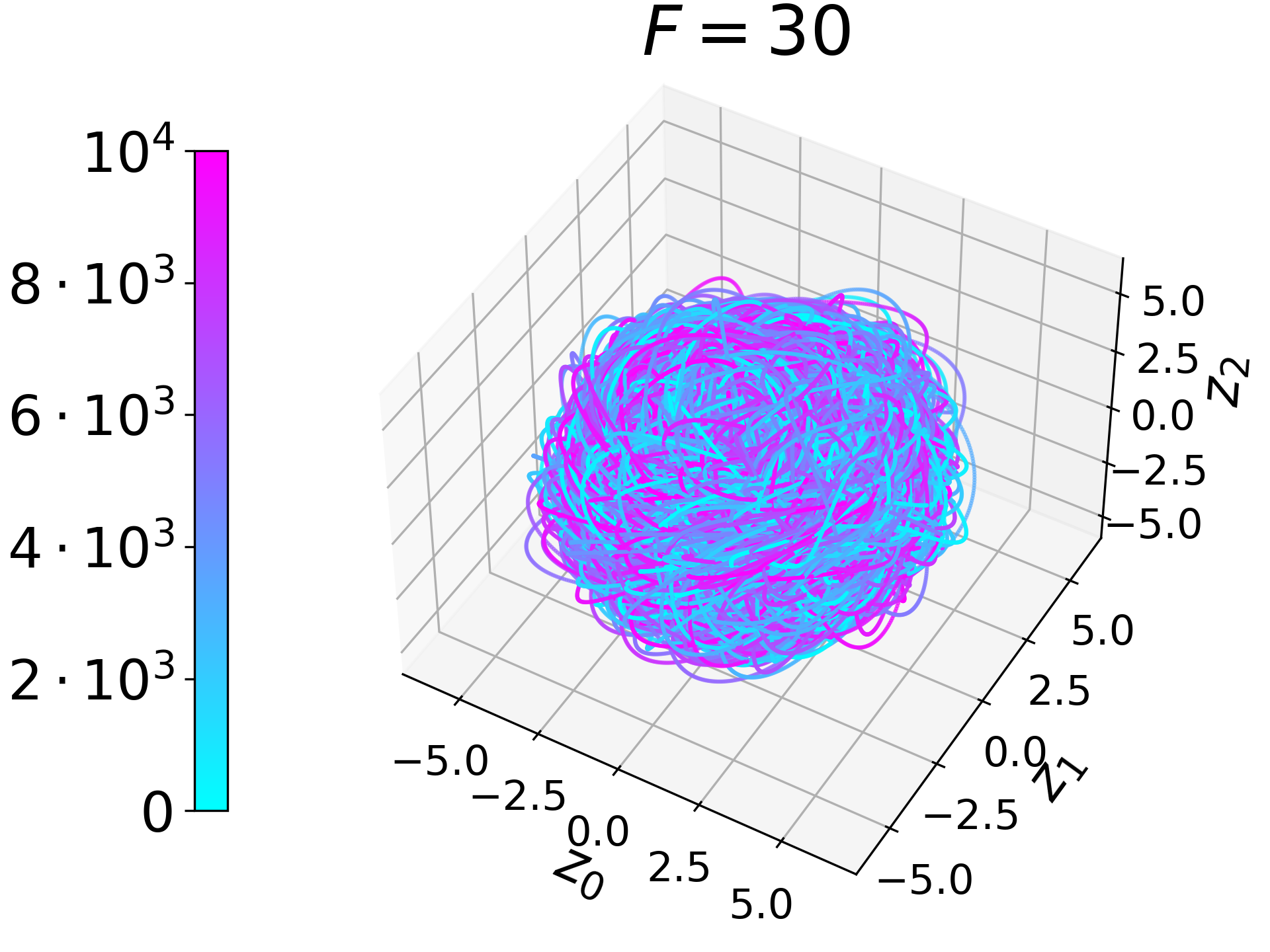}
\caption{Figure illustrates the PCA reconstruction of long trajectories of the L05 II system for $N=30$, $K=3$, and two values of the forcing parameter: $F=5$ (left) and $F=30$ (right). The axes, labeled $z_0$, $z_1$, and $z_2$, correspond to the scores of the first three principal components. These components represent the directions in the original state-space along which the system's variance is the highest. Specifically, they account for $60\%$ of the total variance in the case of $F=5$, and $30\%$ when $F=30$. Both systems are simulated for $10^4$ time units, as indicated on the colorbar, starting from a random point in the state-space, generated in accordance with Lorenz\cite{L05}.}
\label{attractor-pca}
\end{figure}

The principal component analysis\cite{gibson1992analytic} (PCA) of the L05 II's long trajectories, which serve to represent the system's attractor with parameters $N=30$ and $K=3$, is displayed in Fig. \ref{attractor-pca} for two distinct forcing term $F$ values. At $F=5$, the system exhibits more regular dynamics, effectively evolving on a manifold much lower than $N=30$. However, the system at $F=30$ is markedly different, lacking a distinct structure due to its highly chaotic dynamics. Despite this chaotic characteristic, the system's state evolution remains tractable when visualizing the time-varying component values side-by-side.

\begin{figure}[!h]
\centering
\includegraphics[width=0.49\textwidth]{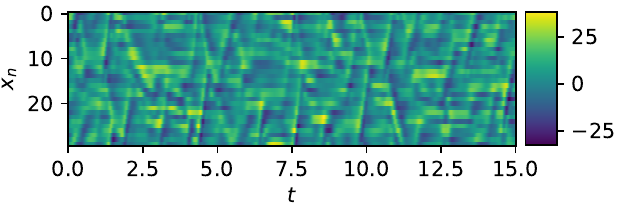}
\caption{Typical time evolution for L05 II at the value of model-specific parameters $N=30$, $K=3$ and $F=30$. Every row represents the time evolution of one component of the state vector$\mathbf{x}(t)$ over time range $[0,15]$.}
\label{heatmap}
\end{figure}

In this vein, Fig. \ref{heatmap} depicts a typical trajectory of the L05 II system at $N=30$, $K=3$, and $F=30$ in the form of a heat map. Each component of the state vector $\mathbf{x}(t)$ is constrained within a finite interval and locally resembles a superposition of waves. The aforementioned transport of the scalar variable across the closed chain manifests as conspicuous ridges, slightly deviated from the vertical direction of the plot.

Along the three model-specific parameters that allow for tuning of the system's certain properties, L05 II possesses additional two key features that play an essential role in this study. First, the model is manifestly translation symmetric along the chain, i.e., it is invariant under the cyclic change in variables indices
\begin{equation}\label{translation symm}
n\to n+k\quad (\text{mod}\ N)
\end{equation}

\noindent
for an arbitrary integer $k$.\footnote{'mod' denotes the modulo operator} The STLSQ algorithm does not have a property that would preserve this symmetry on purpose.

Second, if we rewrite the model Eq. \eqref{model2} in a tensorial notation as
\begin{equation}\label{ODE-form}
\frac{d x_{n}}{d t} = f_n = F + \sum_{n_i=0}^{N-1}L_{nn_i}x_{n_i}+\sum_{n_i=0}^{N-1}\sum_{n_j=0}^{N-1}Q_{nn_in_j}x_{n_i}x_{n_j},
\end{equation}
we find that the tensor $Q_{nn_in_j}$ is skew-symmetric and that $L_{nn_i}$ is negative definite. The long-term stability theorem \cite{Onlong-termboundednessofGalerkinmodels} guarantees for systems that possess this property that there exists a $N$-dimensional ellipsoid in the state-space into which the trajectory resulting from the general initial condition will fall after some initial finite trapping time and remain there forever. Since such systems, therefore, do not generate divergent trajectories, we can assume that the long-term dynamical and statistical properties that will be presented in Sec. \ref{ch evaluation methods models} are reasonably defined for the case of L05 II. 

Another feature that comes about as a direct consequence of the model's equation form, is that the divergence of the system's ODE is equal to the negative value of the system's dimension, i.e.,
\begin{equation}\label{divergence}
\sum_n\frac{\partial f_n}{\partial x_n}=-N.
\end{equation}

\subsection{Utilized evaluation methods of estimated models}\label{ch evaluation methods models}

In this section, we will discuss the methods of dynamical and statistical analysis that were selected as suitable characteristics on basis of which the source system and the estimated models can be compared. Specifically, the evaluation of the estimated models will be based on the deviation from the source system in terms of the Lyapunov spectrum, power spectral density (PSD) and an example-specific function that measures the breaking of a certain symmetry that is present in the source system. Following in the Sec. \ref{ch evaluation methods estimator}, we will separately discuss
the evaluation of the estimator in terms of the covariant matrix of estimated model coefficients, which will provide insights into the sensitivity of the STLSQ algorithm to variations in data.

\subsubsection{Lyapunov exponents}\label{ch lyap}

Suppose that a $N$-dimensional dynamical system under inspection possesses $N$ Lyapunov exponents \cite{Ott} that can be numerically calculated, e.g. as in our case, via standard Benettin algorithm\cite{Benettin} with modified Gram-Schmidt orthonormalization\cite{trefethen1997numerical}. Lyapunov exponents dictate the rate and the fashion in which the control volume of the dynamical system's state-space will deform under the action of time evolution in the statistical limit. Particularly, its largest value determines the system's sensitivity to perturbations in initial conditions. 

Further on, the spectrum as a whole can be used to approximately determine Lyapunov dimension\cite{KaplanYorke} $d_L$ that serves as a measure of the dimension of the attractor's manifold, i.e., the subspace of the state-space, on which the systems' dynamics effectively takes place. Suppose that we order the exponents $\lambda_i$ in the spectrum from the most positive $\lambda_1=\lambda_{max}$ to the most negative $\lambda_N=\lambda_{min}$. The Lyapunov dimension can be calculated using the Kaplan-Yorke formula\cite{KaplanYorke}:
\begin{equation}\label{dL}
d_L=j+\frac{\sum_{i=1}^{j}\lambda_i}{|\lambda_{j+1}|},
\end{equation}

\noindent
where $j$ is the highest index for which $\sum_{i=1}^{j}\lambda_i>0$.

Another property of the Lyapunov spectrum relevant for our case is the equivalence of the divergence of the system's ODE and the sum of all exponents $\lambda_i$ in the spectrum \cite{sandri1996numerical}, i.e.,
\begin{equation}\label{divergence2}
\sum_n\frac{\partial f_n}{\partial x_n}=\sum_i \lambda_i.
\end{equation}

If the divergence of the system's ODE turns out to be a state-space constant, as in the case of \eqref{divergence}, it can be used as a numerical check when computing the Lyapunov spectrum numerically.

The Lyapunov spectrum is known to be invariant under smooth changes of the coordinate system\cite{parlitz1993lyapunov}. Consequently, only alterations in the estimated model parameter space that contribute to the change in its dynamics will lead to variations in the Lyapunov spectrum values and other closely related quantities. With this in mind, we regard the Lyapunov spectrum and other closely related quantities as valuable characteristics for evaluating estimated models.

Lyapunov exponents govern the behavior of the distance between initially adjacent points in state-space under the system's time evolution. For dynamical systems with finite-size attractors, this is, however, relevant only for short time periods. To encapsulate the system's long-term properties, it is sensible to examine a different quantity. As a natural candidate presents itself the power spectral density\cite{horvatsirca}.

\subsubsection{Power spectral density}\label{ch PSD}

Power spectral density (PSD) holds the information on the presence of waves of different frequencies $\nu$ within the signal $\mathbf{x}_n$, where an instance of the signal is the time series generated by sampling system's $n$th component $x_n$ with a discrete time step $\Delta t$ over a finite time interval $t\in[0,(m-1)\Delta t]$, where the number of samples is $m$. In other words, $\mathbf{x}_n$ is a column vector in the matrix of system states \eqref{matrix of the system states}.

Such a signal, generated by a chaotic dynamical system, will result in a PSD that is in general a highly variable function and must be thus appropriately smoothed in order to be compared between source system and estimated models. For this purpose, we employ a well-known method for smoothing the noise in the spectrum; the so-called Welch's method \cite{Welch}. The source dynamical system possesses a translation symmetry \eqref{translation symm}, but this only holds approximately for estimated models. Consequently, it is sensible to get additional smoothing of the spectra by averaging over all $n=0,\dots,N-1$ signals $\mathbf{x}_n$ produced by the corresponding system components. That is, we define our observable to be 
\begin{equation}\label{S-tilde}
\tilde{S}(\nu)=\frac{1}{N} \sum_{n=0}^{N-1} S(\nu| \mathbf{x}_n),
\end{equation}
where $S(\nu| \mathbf{x}_n)$ denotes the PSD of the signal at frequency $\nu$ generated by the $n$th component of the system, calculated via Welch's method. The parameters of Welch's method were adjusted to produce PSDs for trajectories of the source system and different estimated models, that were, at an adequate frequency resolution, sufficiently smooth to be compared with each other.

\subsubsection{Spatial correlation}\label{ch spatial correlation}

We want to design a quantity to describe the interdependence of the components of the dynamical system's ODE, i.e., its state-space variables. Let us define the spatial correlation matrix $\mathbf{C}$ of dimension $N\times N$ in terms of its coefficients:
\begin{equation}\label{Cij}
C_{n_1,n_2} = \frac{1}{m} \sum_{i=0}^{m-1} (x_{n_1,i} - \bar{x}_{n_1,i}) (x_{n_2,i} - \bar{x}_{n_2,i}).
\end{equation}

\noindent
As in Sec. \ref{ch PSD}, $\mathbf{x}_{n_1}$ is a signal corresponding to the system’s $n_1$th component, specifically $x_{n_1,i}$ is its $i$th entry and $\bar{x}_{n_1,i}$ denotes the mean of each signal $\mathbf{x}_{n_1}$, i.e., $\bar{x}_{n_1,i}=\frac{1}{m}\sum_i x_{n_1,i}$. To simplify the notation in the following equations we define the matrix $\mathbf{C}$ to be cyclically periodic in each index with period $N$, i.e.,
\begin{equation}
    C_{n_1+N,n_2} = C_{n_1,n_2+N} = C_{n_1,n_2}.
\end{equation}

Matrix $\mathbf{C}$ is manifestly symmetric (under exchange of indices) and contains information about the correlation between two arbitrary components of the system's state vector. Its diagonal values ($n_1$=$n_2$) are a well-known quantity, i.e., the variance of the values within the signal $\mathbf{x}_{n_1}$, produced by $n_1$th system component. In the case that, cyclically relabeling state-space variable indices does not change the equation of motion \eqref{ODE}, i.e., the system has symmetry \eqref{translation symm}, for any integer $k$ it also holds 
\begin{equation}\label{cyclic-symmetry}
C_{n_1,n_2}=C_{n_1+k,n_2+k}.
\end{equation}

These symmetries reduce $N \times N$ of initially independent quantities of the spatial correlation matrix $\mathbf{C}$ to just $N/2+1$ if $N$ is even or $(N+1)/2$ if $N$ is odd. For this purpose, it is convenient to introduce
\begin{equation}\label{C1k}
C^{(1)}_k = \frac{1}{N}\sum_{n=0}^{N-1} C_{n, n+k},
\end{equation}
\noindent
where the integer $k$ runs from 0 to $N/2$ or to $(N-1)/2$.

In fact, symmetry \eqref{cyclic-symmetry} is expected to hold in the limit of the infinite continuous signal, i.e., when we send the number of samples $m$ and the sampling rate $1/\Delta t$ to infinity. However, we expect the symmetry to approximately hold for sufficiently long signals. Estimated models, or specifically their ODEs in general will not possess symmetry \eqref{cyclic-symmetry}. Deviation from it can be characterized by
\begin{equation}\label{C2k}
C^{(2)}_k = \frac{1}{N}\sum_{n=0}^{N-1} (C_{n, n+k}-C^{(1)}_k)^2,
\end{equation}
which can be thought of as the variance of $C_{n,n+k}$ at fixed $k$. The violation of the symmetry \eqref{cyclic-symmetry}, which is likely to be observed in estimated models, will be reflected in larger values of $C^{(2)}_k$ compared to the source system.

\subsection{Evaluation of the estimator: Covariance matrix of model coefficients}\label{ch evaluation methods estimator}

Regardless of our knowledge of the source system, it is possible to statistically evaluate the estimator based on the sensitivity (or dispersion) of the model coefficients in the presence of small changes in the data. That is, the sensitivity of the solution $\bm{\Xi}$ of the Eq. \eqref{problem eq} found by the investigated algorithm at some specific value of the data $\mathbf{X}$ is described by the covariance matrix of model coefficients\cite{seber2005nonlinear}.

Let us express the matrix of model coefficients $\bm{\Xi}$ and the matrix of data $\mathbf{X}$ in a vectorized form, denoted by $\bm{\Xi}_V$ and $\mathbf{X}_V$. For sufficiently small perturbation of the data $\mathbf{X}_V\to\mathbf{X}_V+\delta\mathbf{X}_V$ the variation in $\bm{\Xi}_V$ is linearly dependent on $\delta\mathbf{X}_V$, i.e.,
\begin{equation}\label{lin-pert}
\delta \bm{\Xi}_V  = \mathbf{D} \delta \mathbf{X}_V + \mathcal{O}(\delta \mathbf{X}_V^2),
\end{equation}
for some matrix $\mathbf{D}$ containing information about the model and minimization procedure.

Suppose the perturbation $\delta\mathbf{X}_V$ is originating from white Gaussian noise $\mathcal{N}(0,b^2)$ with zero mean and standard deviation $b$ in the data. The model sensitivity to such variations of the data is characterized by the covariance matrix of model coefficients $\rm{Cov}[\delta\text{$\bm{\Xi}_V$}]\in\mathbb{R}^{Q\times Q}$ with $Q=PN$, i.e., 
\begin{equation}\label{cov-eq-teo2}
\rm{Cov}[\delta\text{$\bm{\Xi}_V$}]=\mathbf{D}\rm{Cov}[\delta\mathbf{X}_V]\mathbf{D}^T=\mathbf{D}\mathbf{D}^T \textit{b}^2,
\end{equation}

\noindent
where we took into account the manifestly diagonal form of data covariance matrix $\rm{Cov}[\delta\mathbf{X}_V]=\mathbf{I}\textit{b}^2$.

The covariance matrix can be approximated well by iteratively fitting the estimated model to data for several, say $M$ random $(\delta\mathbf{X})_j$ around a fixed $\mathbf{X}$ and from the resulting $(\delta\bm{\Xi})_j$ evaluating the sample covariance matrix \cite{sample-cov}
\begin{equation}\label{cov-eq-emp}
\rm{Cov}[\delta\text{$\bm{\Xi}_V$}]=\frac{1}{M-1}\sum_{j=0}^{M-1}(\delta\text{$\bm{\Xi}_V$})_j(\delta\text{$\bm{\Xi}_V$})_j^T. 
\end{equation}

The corresponding correlation matrix elements $(\rm{Corr}[\delta\bm{\Xi}_V])_{q_1,q_2}$, $q_1,q_2\in\{0,1,\dots,Q-1\}$, are calculated from the covariance matrix as

\begin{equation}\label{corr-eq}
\begin{aligned}
    &(\rm{Corr}[\delta\bm{\Xi}_V])_{q_1,q_2}=\\
    &(\rm{Cov}[\delta\bm{\Xi}_V])_{q_1,q_2}\big((\rm{Cov}[\delta\bm{\Xi}_V])_{q_1,q_1}(\rm{Cov}[\delta\bm{\Xi}_V])_{q_2,q_2}\big)^{-\frac{1}{2}}.
\end{aligned}
\end{equation}

\section{DATA GENERATION AND WORKFLOW}\label{ch workflow}

In order to provide context and clarity for the reader, the current section outlines the workflow that was carried out and that we propose along with the key details of our study. In short, we first selected a specific model instance (i.e., L05 II at selected parameters $N$, $K$ and $F$) and simulated the model noise-free to obtain time series (i.e., discrete points along trajectories) on which we evaluated certain characteristic functions. Subsequently, we added white Gaussian noise to the L05 II's time series, which can be viewed as introducing measurement noise\cite{schreiber1996observing}, and then applied the STLSQ algorithm to this modified data. The same characteristic functions were then evaluated on the resulting estimated models generated from the noisy data. The entire process is described in detail below:

\begin{enumerate}
    \item 
    First, we chose our specific model to be the L05 II at the value of model-specific parameters
    \begin{equation}\label{NKF}
    N=30,\quad K=3,\quad F=30.
    \end{equation}
    
    The specific dimension parameter $N=30$ along with the forcing parameter $F=30$ was carefully chosen so that the dynamics of the system is complex enough (chaotic) and its attractor is of higher dimensionality than in the examples discussed in previous studies \cite{SINDy}\textsuperscript{,}\cite{pysindy}. On the other hand, it was also desirable to keep the computational complexity sufficiently low for the program to be run on a personal computer.
    
    The model-specific parameter $K=3$ gave rise to 18 bilinear terms on the RHS in every component, i.e., 20 terms in total. We chose a polynomial library $\bm{\Xi}$ of order $2$, i.e., the space of all possible function terms to be the space of all constant, linear and bilinear terms, that equaled to $P\cdot N=14880$ free parameters of the estimated model. Thus, the vector representing components of the source model's ODE has an adequately sparse representation in this basis, supporting the use of a sparse regression technique STLSQ. The algorithm did not produce satisfactory results for polynomial libraries of higher orders.
    
    The model-specific parameters \eqref{NKF} were also chosen such that the amplitudes of the coefficients in the ODE range from roughly $\mathcal{O}(10^{-1})$ to $\mathcal{O}(10^{1})$ posing another challenge for the algorithm.
    
    \item\label{step2}
    With a specific model in hands, we generated a random initial condition as described in Lorenz's paper\cite{L05} and propagated it forward $10^3$ time units to reach the attractor, where we assumed the dynamics to be sufficiently ergodic and thus the Lyapunov exponents to be meaningfully defined. Furthermore, we supposed the existence of a global attractor, a premise that was employed also by other researchers of the L05 models\cite{L05}\textsuperscript{,} \cite{LE98-birfu}. To support this premise, we emphasize that different time averages (such as moments, Lyapunov exponents, etc.) turned out to be independent of initial conditions, a result that would in general not be expected in the presence of multiple attractors. Moreover, the L05 II satisfies the criteria of the long-term stability theorem \cite{Onlong-termboundednessofGalerkinmodels}, and as a consequence has a region of attraction clearly defined.

    Once we obtained a new initial condition situated on the attractor, we propagated it further in time for an additional $10^4$ time units, sampling the trajectory at every $\Delta t=10^{-3}$, and storing the data in matrix $\mathbf{X}_0$. Using an ordinary Fourier Transform (FT), we first estimated the Nyquist-Shannon critical sampling frequency \cite{shannon} above which information loss becomes negligible. This step is vital, as previous studies have emphasized that undersampling can result in estimated models with less developed dynamics\cite{letellier2009frequently}. Then, we employed Welch's method to obtain a smoothed PSD spectrum of the source system. On that very same trajectory, we also determined the values of $C^{(1)}$ and $C^{(2)}$ for the source system.
    
    Next, we calculated the Lyapunov spectrum (with the initial condition being the first point on the saved trajectory $\mathbf{X}_0$) and tuned the parameters of the Benettin algorithm to values where condition \eqref{divergence} was well approximated. The Benettin algorithm parameters remained fixed for all future calculations. 
    
    To reduce the possible bias (deviation from the theoretical result in the statistical limit) presented by random initial conditions, the whole procedure was repeated multiple times; the results presented in the next chapter are averaged over five different random initial conditions and corresponding trajectories $\mathbf{X}_0^{(\beta)}$ for $\beta=1,\dots, 5$. Additionally, averaging over 5 samples was enough to produce to considerably smooth PSDs.
    
    \item\label{step3}
    In the next step, we first selected one of the saved trajectories, say $\mathbf{X}_0^{(\beta)}$, as the data of the STLSQ algorithm. Specifically, the data were composed of the initial $m$ samples from the trajectory, selecting only every tenth data point. In other words, we pruned the data by retaining only every tenth row of $\mathbf{X}_0^{(\beta)}$. This approach effectively set our sampling frequency to $\Delta t=10^{-2}$, a value determined in accordance with the Nyquist-Shannon critical sampling frequency. Before proceeding with executing the STLSQ algorithm, white Gaussian noise $\mathcal{N}(0,a^2)$ with mean $0$ and standard deviation $a$ was added to the data, i.e., to source system's time series.\footnote{To each matrix element of $\mathbf{X}_0^{(\beta)}$ we added a random number from $\mathcal{N}(0,a^2)$.} For the purpose of our current discussion, we will denote the data that includes measurement noise with standard deviation $a$ as $\mathbf{X}_{a}^{(\beta)}$.   
    The optimal number of supplied samples $m$ and the algorithm-intrinsic parameters (such as threshold $\lambda$ and regularization intensity $\alpha$)\footnote{Cross-validated algorithm-intrinsic parameters: $\lambda\in\{10^{-3},10^{-2},10^{-1},$ $10^{0}\}$ and $\alpha\in\{0,10^{-3},10^{-2},10^{-1},$ $10^{0}\}$. The lowest threshold value was set to the level at which the fit on noise-free data still produced a sparse ODE. On the other hand, the highest threshold value was chosen just above the size of the smallest coefficient terms in the ODE of the source model. This was done because, at this threshold, the fitted models were no longer capable of accurately capturing the dynamics of the source model.} were chosen through cross-validation, suitable for the temporal nature of the data.\footnote{The STLSQ algorithm needs to differentiate the data, we need the training and test data to consist of sequential time intervals. \cite{pysindy}} Specifically, we used \textit{scikit}'s function \textit{TimeSeriesSplit}\footnote{At default parameters (5 splits) and a number of samples $m=2\cdot 10^4$.}\textsuperscript{,}\cite{scikit} that is a variation of $k$-fold which returns first $k$ folds as train set and the $(k+1)$th fold as test set. The estimated model fitted on $\mathbf{X}_{a}^{(\beta)}$, which we will denote as ${\rm M_{est}}(a,\beta)$, was then obtained by refitting on all $m$ data samples at the found optimal algorithm-intrinsic parameters.
    
    \item Utilizing the STLSQ algorithm with the most effective intrinsic parameters $\alpha$ and $\lambda$ as determined from non-noisy data $\mathbf{X}^{(\beta)}_0$, we carried out a statistical evaluation of the estimator. In particular, we examined how the bias and dispersion (represented by the covariance matrix) of the estimated models' coefficients varied across different levels of white Gaussian noise standard deviation $b$, as discussed in Sec. \ref{ch evaluation methods estimator}. For each considered perturbation amplitude $b$, we selected one of the saved trajectories, say $\mathbf{X}^{(\beta)}_0$, and ran the STLSQ algorithm $M=100$ times on the $100$ different realizations of $\mathbf{X}^{(\beta)}_b$. Additionally, the results were averaged over all $\beta=1,\dots,5$ saved trajectories. By using $b$ to denote the perturbation amplitude (i.e., small noise level) in this part of the study (instead of $a$), we emphasize the statistical nature of the results from this evaluation of the estimator, differentiating it from the specific example study concerning properties of the estimated models addressed in the subsequent step.
    
    \item Step \ref{step3} was repeated for multiple values of white Gaussian noise standard deviation $a\in\{0,0.5,1.0,1.5\}$ and for all $\beta=1,\dots,5$ different data instances within each noise level, resulting in a total of 20 different estimated models $\rm{M_{est}}$($a$,$\beta$). We emphasize that the optimal values of the threshold and regularization parameters $\lambda$ and $\alpha$ were chosen through cross-validation individually for each data instance. The selected values of noise levels $a$ approximately correspond to $a\in\{0$, $0.0093\sigma_A$, $0.0185\sigma_A$, $0.0277\sigma_A\}$. With $\sigma_A$ we denoted the standard deviation of the $m$ sampled points on the source system's trajectory, i.e.,
    \begin{equation}\label{attr-size}
    \sigma_A = \sqrt{\frac{1}{m}\sum_{n=0}^{N-1}\sum_{i=0}^{m-1}(x_{n,i}-\bar{x}_{n,i})^2}
    \end{equation}

    and can be regarded as the measure for the size of the attractor. The notation in the above equation is consistent with one introduced in Sec. \ref{ch spatial correlation}. The value of $\sigma_A$ was averaged over all five source system's example trajectories $\mathbf{X}_0^{(\beta)}$. 
    
    Further on, for each obtained estimated model $\rm{M_{est}}$($a$, $\beta$) we calculated the selected dynamical and statistical properties of the estimated models, i.e., Lyapunov spectra, PSD, $C^{(1)}$ and $C^{(2)}$. The method-specific and other parameters of the utilized methods (i.e., initial conditions, Benettin algorithm parameters, Welch's method parameters, number of samples, and sampling frequency) were kept the same throughout this procedure. Finally, averaging the results over all five estimated model instances within each noise level $a$, we plotted the graphs representing deviations of estimated models from the source system in terms of all selected properties and for all chosen noise levels $a$. For the sake of brevity, we will refer to the calculations averaged over different estimated model instances $\beta$ within one noise level $a$ as properties belonging to estimated model, ${\rm M_{est}}(a)$.
\end{enumerate}

The results were calculated using program code written in $Python$ programming language (version $3.7.3.$). Specifically, we used \textit{scipy}'s \textit{LSODA} integrator with automatic stiffness detection and switching\cite{petzold1983automatic}\textsuperscript{,}\cite{hindmarsh1983odepack}. The data was represented and managed using \textit{numpy} library \cite{numpy} and the STLSQ algorithm is available on the official \textit{pysindy} repository \cite{pysindy}.

\section{RESULTS AND DISCUSSION}\label{ch results}

In this section, we will present the results of our study. In Sec. \ref{ch res estimator}, we will first report on the bias in the estimated model coefficients resulting from non-noisy data input to the estimator. This will be followed by a discussion on the sensitivity of model coefficients to small noise levels, quantified through the covariance and correlation matrices of the estimated model coefficients. As we shift focus to higher noise levels in Sec. \ref{ch res estimated models}, we will illustrate the influence of the measurement noise in the data with respect to the dynamical and statistical properties of the estimated models, outlined in Sec. \ref{ch evaluation methods models}.

\subsection{Evaluation of the estimator}\label{ch res estimator}

All the models trained on noise-free data [i.e., $\rm{M_{est}}$($a=0$)] turned out to be slightly biased. It is worth noting that in each example the threshold parameter $\lambda$ (see Sec. \ref{ch SINDy}), chosen accordingly to cross-validation results, was two orders of magnitude smaller than the smallest coefficient present in the source system's ODE. Additionally, the regularization strength parameter $\alpha$, which was also determined through cross-validation, was non-zero in all noise-free data cases. Consequently, the estimated model's ODE contain some additional non-zero linear terms which contribute to the bias of model coefficients. 

\begin{figure}[h!]
\centering
\includegraphics[width=.48\textwidth]{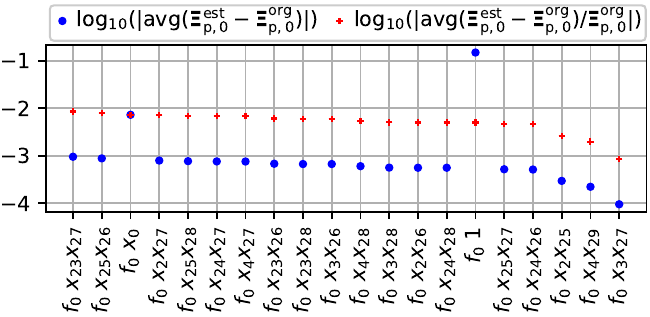}
\caption{Absolute (blue) and relative (red) bias in coefficients of the first component of the $\rm{M_{est}}$($a=0$) estimated model's ODE (i.e., $\boldsymbol{\Xi}_{p,0}:=\boldsymbol{\Xi}^{\rm{est}}_{p,0}$) with respect to true coefficients (ones present in the first component of the source system's ODE (i.e., $\boldsymbol{\Xi}^{\rm{org}}_{p,0}$) that can be deduced from Eq. \eqref{model2}). 
The ticks on the abscissa label the coefficients associated with different terms. First, $f_{\bullet}$ denotes the estimated model's ODE component with the index '$\bullet$'. Second, '1' denotes the constant term, "$x_{\bullet}$" represents linear term with index '$\bullet$', and "$x_{\bullet}\ x_{\bullet}$" are the bilinear terms.}
\label{err}
\end{figure}

Figure \ref{err} shows the average and relative bias in estimated model $\rm{M_{est}}$($a=0$) coefficients corresponding to function terms that were present in the first component of source model's ODE. The absolute bias is the greatest in the coefficient corresponding to the constant term. Relative biases of estimated model coefficients (i.e., bias normalized with coefficient values of the source model) are comparable among all estimated model coefficients and are approximately of order $10^{-2}$.

\begin{figure}[!h]
\centering
\includegraphics[width=.47\textwidth]{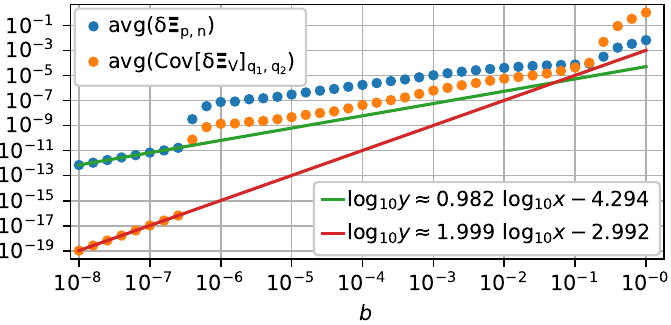}
\caption{The dependence of the average element size of the matrix of noise-induced bias of the estimated model coefficients $\rm{avg}(\delta\bm{\Xi}_{p,n})$ and the average element size of the covariance matrix of the estimated model coefficients $\rm{avg}(\rm{Cov}[\delta\text{$\bm{\Xi}_V$}]_{q_1,q_2})$, with respect to the perturbation amplitude, i.e., standard deviation of white Gaussian noise $b$. For sufficiently small perturbation amplitudes $b$, the bias size shows a nearly linear trend (green line), while the covariance size scales quadratically with $b$ (red line).}
\label{bias cov avr}
\end{figure}

Studying the effect of white Gaussian noise in the data let us first examine how the perturbation amplitude (i.e., noise level $b\ll1$) affects the noise-induced average bias in estimated model coefficients $\rm{M_{est}}$($b$) (see Sec. \ref{ch evaluation methods estimator} and step 4 in Sec. \ref{ch workflow}). The results are depicted in Fig. \ref{bias cov avr}. In the limit of small perturbation of the data, the average (perturbation induced) bias in the estimated models' coefficients $\bm{\Xi}$ exhibits a linear dependence on the perturbation amplitude $b$. We also see in Fig. \ref{bias cov avr} that in this limit, the covariance matrix scales quadratically with the perturbation amplitude, a result that is in agreement with Eq. \eqref{cov-eq-teo2}. 

\begin{figure}[!h]
\centering
\includegraphics[width=0.48\textwidth]{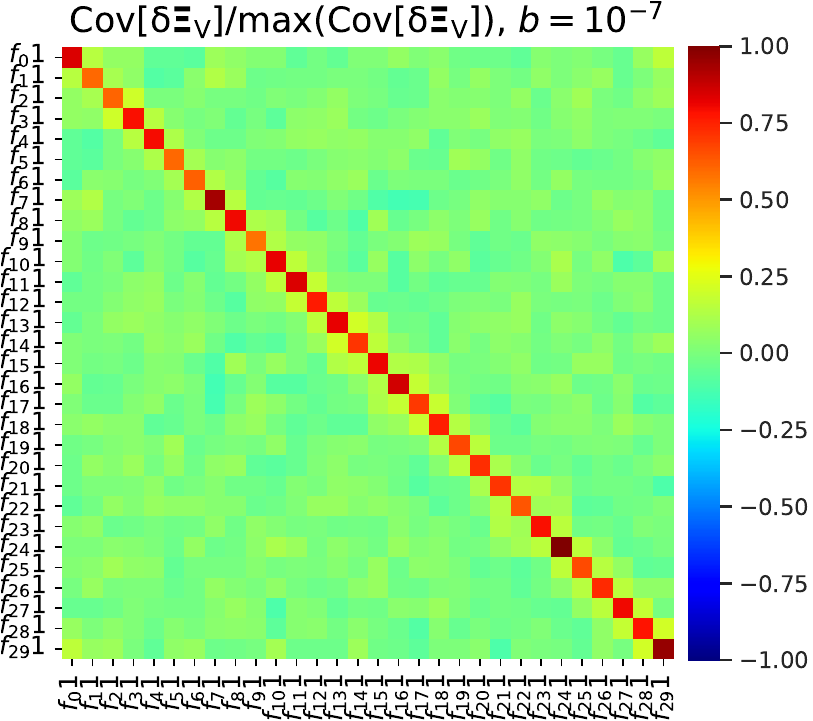}
\caption{A part of the covariance matrix of model coefficients $\rm{Cov}[\delta\text{$\bm{\Xi}_V$}]$ normalized with its maximum element $\rm{max}(\rm{Cov}[\delta\text{$\bm{\Xi}_V$}])\approx4.68\cdot10^{-16}$ for the estimated models trained on non-noisy data. The covariance matrix was calculated following Eq. \eqref{cov-eq-emp} for $j=0,\cdots,M-1=99$, with each $(\delta\bm{\Xi}_V)_j$ being a sample of white Gaussian noise with standard deviation $b=10^{-7}$. The result is averaged over five different estimated model instances, i.e., $\rm{M_{est}}$($a=0$) (see Sec. \ref{ch workflow}). Only columns and rows which contain elements larger than $0.02\cdot\rm{max}(\rm{Cov}[\delta\text{$\bm{\Xi}_V$}])$ are shown. Please refer to Fig. \ref{err} for the notation used in labeling the elements of the covariance matrix.}
\label{cov}
\end{figure}

The covariance matrices of model coefficients $\rm{Cov}[\delta\text{$\bm{\Xi}_V$}]$ [calculated via Eq. \eqref{cov-eq-emp}] are too large (matrices of size $14880\times14880$) to be graphically displayed in full. Thus, only a part of the covariance matrix containing the elements larger than 0.02 of its maximal element are shown in Fig. \ref{cov} for the case of perturbation amplitude $b=10^{-7}$. 
In this perturbation regime, the covariance matrix is approximately diagonal with largest entries corresponding [as for the case of the absolute bias (Fig. \ref{err})] to the constant coefficient terms, i.e., they are at least two orders of magnitude greater than the coefficients belonging to linear and bilinear terms. 

\begin{figure}[!h]
\centering
\includegraphics[width=0.48\textwidth]{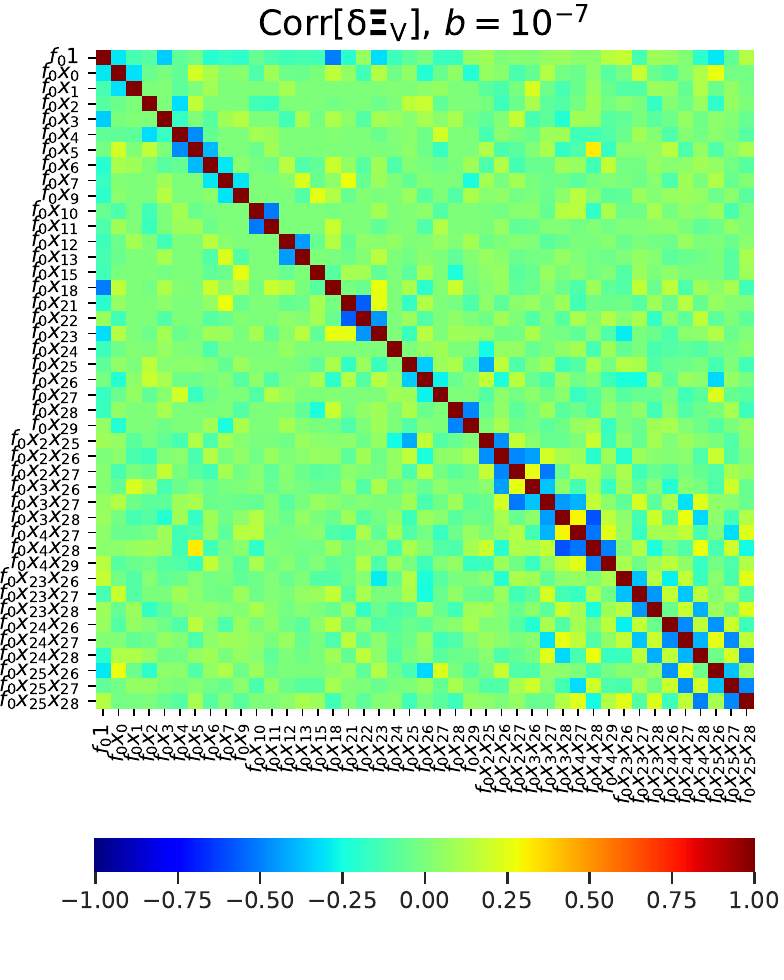}
\caption{The correlation matrix $\rm{Corr}[\delta\bm{\Xi}_V]$ of the estimated model coefficients calculated from the covariance matrix at $b=10^{-7}$ that is partially depicted in Fig. \ref{cov}. Only the non-zero elements corresponding to the first component of the estimated model's ODE are shown. Please refer to Fig. \ref{err} for the notation used in labeling the elements of the correlation matrix.}
\label{corr}
\end{figure}

The correlation matrix at $b=10^{-7}$ indicates that there is a higher correlation among function terms belonging to the same component of the estimated model's ODE. These terms are mostly spatially adjacent linear terms and bilinear terms of type $x_{n_1}x_{n_2}$ and $x_{n_3}x_{n_4}$, where $|n_1-n_3|\leq3$ and $|n_2-n_4|\leq3$, as can be seen explicitly in Fig. \ref{corr}, where we displayed the part of the correlation matrix corresponding to the first ODE component $f_0$. Considerable correlation is also present between function terms of neighboring ODE components, i.e., between terms belonging to ODE components $f_{n_1}$ and $f_{n_2}$ with $|n_1-n_2|=1$. In short, the largest entries of the correlation matrix primarily fall into blocks on the diagonal, corresponding to the same or neighboring components of the estimated model's ODE. This result is influenced by the structure of the source model L05 II, particularly its correlation property between neighboring components of the source system's state-space trajectory (as seen in Fig.e \ref{heatmap}).

\begin{figure}[!h]
\centering
\includegraphics[width=0.48\textwidth]{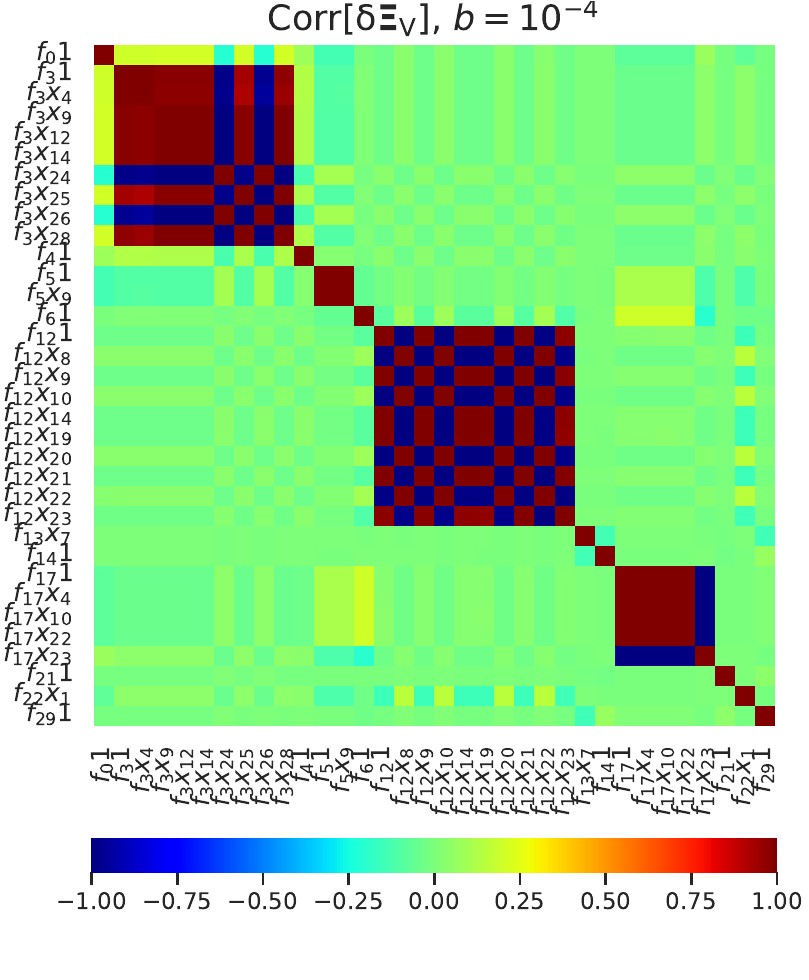}
\caption{The part of the correlation matrix that corresponds to the elements of the covariance matrix at perturbation amplitude $b=10^{-4}$, which are larger than $0.02\cdot\rm{max}(\rm{Cov}[\delta\text{$\bm{\Xi}_V$}])$. Please refer to Fig. \ref{err} for the notation used in labeling the elements of the correlation matrix.}
\label{corr2}
\end{figure}

Returning back to the Fig. \ref{bias cov avr} we see that the linear and quadratic trend of bias and covariance amplitude breaks at a relatively small perturbation amplitude (roughly below $10^{-6}$). Studying the covariance matrices slightly above this threshold, one finds that a few blocks on the diagonal, corresponding to the coefficients associated with terms within a single ODE component $f_n$, suddenly become strongly correlated. As we slowly increase the perturbation amplitude $b$ toward $10^{-6}$ and above more and more of the model coefficients become strongly correlated, that is, with coefficients belonging to the same ODE component. A representative example of the correlation matrix in this regime is shown in Fig. \ref{corr2} for the case of $b=10^{-4}$. As soon as all $N=30$ blocks on the diagonal, corresponding to each ODE component, assume values close to 1 or -1, the correlations slowly start to decrease until reaching the form comparable to the case of $b=10^{-7}$ that is shown in Fig. \ref{corr}. We also noticed an abrupt jump in the ratio between the maximal and average covariance matrix element size at the perturbation amplitude of about $b=1.25\cdot10^{-7}$. Around this value of $b$, the ratio suddenly increases by more than one order of magnitude. As the perturbation amplitude $b$ increases further, the ratio slowly decreases in size until, at $b\approx10^{-1}$, reaching a value comparable to one at $b=10^{-7}$. A very similar trend can also be seen in the ratio between maximal and average coefficient bias size. The reason for this behavior is yet to be understood and will require further study in future work. The second structural change in the trend of the average coefficient matrix and covariance matrix element size (as seen in Fig. \ref{bias cov avr}) occurs just above the value of $b=10^{-1}$. This indicates that at higher noise values, the STLSQ algorithm converges to a different set of function terms in the estimated ODE. Here, changes in the model structure are counterbalanced, to an extent, by alterations in the estimated parameters. Yet, it is worth noting that the fundamental concern lies within the dynamics produced by the final model. In this context, the estimated models trained on noisy data $\mathbf{X}_{a}^{(\beta)}$ with $a\in{0.0093\sigma_A,\ 0.0185\sigma_A,\ 0.0277\sigma_A}$, which are to be discussed in more detail in Sec. \ref{ch res estimated models}, all reside deep within this altered model structure regime.

\subsection{Evaluation of the estimated models}\label{ch res estimated models}

\begin{figure}[!h]
\centering
\includegraphics[width=0.47\textwidth]{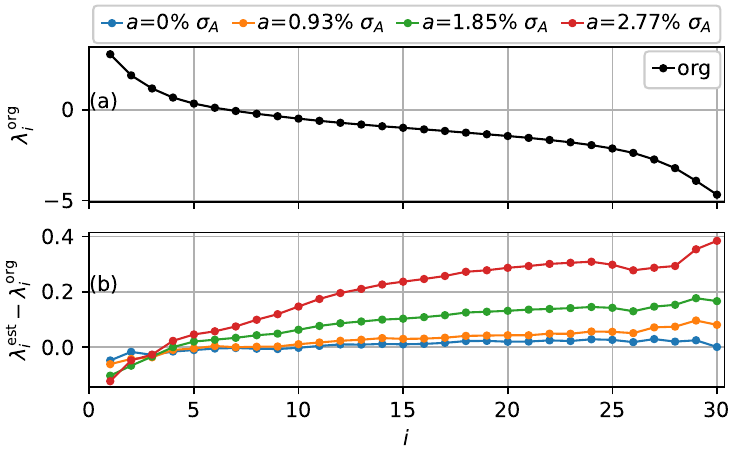}
\caption{(a) Lyapunov spectrum for the L05 II at parameters \eqref{NKF}. (b) The deviations of the Lyapunov spectra of the estimated models $\rm{M_{est}}$($a$) from the source system. Estimated models were obtained from noisy L05 II time series with white Gaussian noise of levels of $a\in\{0,0.0093\sigma_A,0.0185\sigma_A,0.0277\sigma_A\}$.}
\label{lyap-spec}
\end{figure}

The Lyapunov spectrum of the source system is depicted in Fig. \ref{lyap-spec}(a) and has its maximum value at $\lambda_{\rm max}\approx3.08$. Selected Benettin algorithm parameters were: initial perturbation size $\delta_{\rm start}$ was set to $10^{-7}$, the propagation time before each renormalization was 2 time units and the trajectories were renormalized $10^3$ times. For this set of Benettin algorithm parameters, the calculated sum of all exponents in the source system's spectrum was roughly equal to the negative of the dimension of the source system's ODE, i.e., $-N$ (see table \ref{table1}), a result that is in agreement with Eqs. \eqref{divergence} and \eqref{divergence2}.

The Lyapunov spectra of the estimated models are shown in Figure \ref{lyap-spec}(b). First of all, we observe a deformation of the Lyapunov spectrum even for an estimated model fitted on non-noisy data. This observation could be linked to the bias in the estimated model's coefficients that is displayed in Fig. \ref{err}. Higher noise levels, where the bias of the estimated model's coefficients is even larger, result in even more pronounced gradual uniform deformation of the Lyapunov spectrum, and in effect, a greater change in the relevant tracked quantities, i.e., $\lambda_{\rm max}$, $d_L$ and $\sum_i\lambda_i$. However, we are in no position to assert the direct effect of the bias in estimated model coefficients on the model's dynamics as it is well known that correlation does necessarily imply causation. Taking a closer look, the Lyapunov spectra of all estimated models have smaller values of $\lambda_{\rm max}$; moreover, the majority of the exponents tend to lie closer to 0, i.e., we observe a decrease in $|\lambda_i|$ for the vast majority of exponents in the spectrum of each estimated model. This property becomes pronounced for greater noise levels $a$ in the data as can be seen in Fig. \ref{lyap-spec}(b).

\begin{table}[h]
\centering
\begin{tabular}{c|c|c|c|c}
& $\lambda_{\rm{max}}$ & $\sum_i \lambda_i$ & $d_L$ & $\frac{\int \tilde{S}^{\rm{est}}(\nu)\mathrm{d}\nu}{\int \tilde{S}^{\rm{org}}(\nu)\mathrm{d}\nu}$ \\
\hline
$\rm{M_{org}}$ & 3.082 & -29.99 & 17.03 & 1.000 \\
$\rm{M_{est}}$($a=0$) & 3.027 & -29.79 & 16.96 & 1.000 \\
$\rm{M_{est}}$($a=0.93\%\ \sigma_A$) & 3.021 & -29.18 & 17.07 & 1.000 \\
$\rm{M_{est}}$($a=1.85\%\ \sigma_A$) & 2.979 & -27.41 & 17.66 & 1.011 \\
$\rm{M_{est}}$($a=2.77\%\ \sigma_A$) & 2.959 & -24.13 & 18.91 & 1.043 \\
\end{tabular}
\caption{A table representing the change in investigated variables between the original (source) system (labeled with $\rm{M_{org}}$) and estimated models $\rm{M_{est}}$($a$), trained on noisy data with noise level $a$.}
\label{table1}
\end{table}

Since the maximal Lyapunov exponent $\lambda_{\rm max}$ directly determines the system's sensitivity to initial conditions, we see that the noise masks an important feature of the source system's dynamics, i.e., the estimated models generated from noisier data will behave less chaotically. This observation aligns with established understandings in the modeling of chaotic systems, as discussed by Schreiber and Kantz\cite{schreiber1996observing}. In the case, the estimated models would be used to simulate real data, e.g. as meteorological models (L05 II is in essence designed to capture some of the basic properties of general meteorological models) to generate ensemble forecasts, the lower sensitivity to initial conditions of the estimated model may lead to an overestimation of the forecast reliability. 

The increase in the sum of the exponents that is mainly due to the decrease of the absolute values of the negative part of the Lyapunov spectrum suggests a fundamental change of the system’s dynamics in two following ways. First, as a consequence, an arbitrary control state-space volume somewhere in the vicinity of the attractor will be pushed toward the local stable manifold more slowly. In effect, one intuitively expects the increase in the fractal dimension as the results on Lyapunov fractal dimension $d_L$ confirm. Second, the deviation of the sum $\sum_i\lambda_i$ from the value of the minus of the system's dimension $N$ as discussed in Secs. \ref{ch L05} and \ref{ch lyap} [equations \eqref{divergence} and \eqref{divergence2}] suggests a gradual departure from the L05 II's ODE form, specifically from the skew-symmetry of the tensor of the quadratic terms $Q_{nn_in_j}$ and the negative definiteness of matrix $L_{nn_i}$. As a consequence, for a sufficiently large level of standard deviation of white Gaussian noise added to the L05 II time series, the estimated models lose an essential property of L05 II, i.e., the property of having a global region of attraction\cite{Onlong-termboundednessofGalerkinmodels}.

\begin{figure}[!h]
\centering
\includegraphics[width=0.47\textwidth]{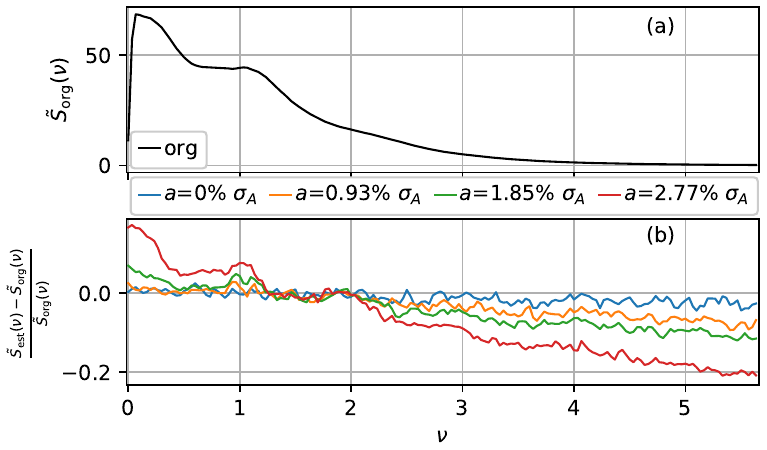}
\caption{(a) Power spectral density (PSD) for the L05 II at parameters \eqref{NKF}, averaged over all system's components, i.e., $\tilde{S}(\nu)$ (equation \eqref{S-tilde}). (b) Relative deviations of $\tilde{S}(\nu)$ of the estimated models $\rm{M_{est}}$($a$) from the source system. The window function as a parameter of Welch's method was chosen to be the Hanning window with a length of $3\cdot10^{4}$ data points. The overlap between windows was maximal, i.e., 50\%.}
\label{PSD}
\end{figure}

The results showed that the PSDs of estimated models $\rm{M_{est}}$($a$) resemble the PSD of the source system (Fig. \ref{PSD}). However, the oscillations with frequencies that dominated the signal of the source system were even more pronounced in the PSDs of the estimated models. This is not solely due to a shift toward lower frequencies (as can be seen in Fig. \ref{PSD}), as the total power $\int \tilde{S}^{\rm{est}}(\nu)\mathrm{d}\nu$ carried by the waves produced by $\rm{M_{\rm{est}}}$($a\neq0$) also increases with respect to the total power carried by the waves in the source system $\int \tilde{S}^{\rm{org}}(\nu)\mathrm{d}\nu$ (see table \ref{table1}). As a result, the estimated models trained on data with higher noise levels exhibit more regular dynamics, which is accompanied by a decrease in chaoticity as indicated by lower values $\lambda_{\rm{max}}$. On the other hand, frequencies that were absent in the source system are also absent in the estimated models. This is a positive finding, as the presence of frequencies in the estimated models that were not present in the source system would imply the existence of some kind of new physics that the source system does not possess.

Using PSD, calculated with Welch's method, as a way to evaluate the estimated models can be very helpful, particularly when dealing with noisy signals from real-life dynamical systems. Welch's method is highly resistant to noise in the signal, making it a reliable characteristic to consider when searching for an appropriate estimated model.

\begin{figure}[!h]
\centering
\includegraphics[width=0.48\textwidth]{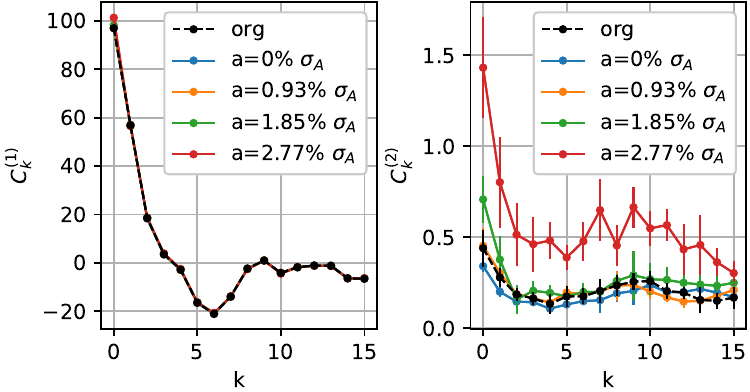}
\caption{Spatial correlation functions $C^{(1)}_k$ (a) and $C^{(2)}_k$ (b) for the source system and estimated models $\rm{M_{est}}$($a$) for different noise levels $a$. The errorbars indicate the standard deviation of the results obtained from multiple simulation iterations.}
\label{spatialcorr}
\end{figure}

Spatial correlation functions $C^{(1)}_k$ and $C^{(2)}_k$ of the source system and estimated models $\rm{M_{est}}$($a$) are shown in Figs. \ref{spatialcorr} (a) and (b). In the first, we observe a minimum at $k=6$ that corresponds to anti-correlation between components with indices $n_1$ and $n_2$ at $|n_1-n_2|=6$. This minimum can be linked to the shape of typical waves (Fig. \ref{heatmap}); the prominent peaks are mostly followed by deep valleys. The position $k=6$ is conditioned by the speed of travel of the wave along the chain of $N$ nodes. 

It is clear that the estimated models approximately retain the form of $C^{(1)}_k$, i.e., the correlation between the system's components at distance $k$ on average does not change much. The same does not hold for $C^{(2)}_k$ (equation \eqref{C2k}), which, as discussed in Sec. \ref{ch evaluation methods models}, represents the variance of the set $C_{n,n+k}$ at fixed $k$ and can be understood as a measure of the violation of the translation symmetry of the source model [and, thus, violation of \eqref{cyclic-symmetry}]. Increasing noise level $a$ in the data, we found that $C^{(2)}_k$ assumes increasingly higher values, indicating the disparities among the components of the estimated model's ODE [Figure \ref{spatialcorr} (b)]. Nevertheless, it is evident that for lower noise levels $a$ the translation symmetry is well respected, at least in the context of $C^{(2)}_k$. 

To conclude, it is evident that for small noise levels in the data, the STLSQ algorithm performs fairly well, that is to say, it mimics the source system, at least in terms of the investigated dynamical and statistical properties. It is about the value of Gaussian noise level $a=1.85\%\ \sigma_A$ where the estimated models start to notably deviate from the source system. The upper value of noise level where the properties of the estimated models were thoroughly inspected was $a=2.77\%\ \sigma_A$. At greater noise levels, such as $a=3.5\%\ \sigma_A$, we found that the STLSQ algorithm only produced estimated models whose trajectories were unbounded, thus, these models were classified as inappropriate. This is to be expected, as our estimator lacks any stringent constraints - the second-order polynomial library provides a flexible framework capable of addressing a multitude of physical problems. Our only true assumption here is the sparsity of the estimated ODEs. Nevertheless, the intriguing question is whether the instability in estimated models primarily arises from the change in model structure (i.e., variations in the set of function terms present in the estimated ODE) or from the poor estimation of the coefficients corresponding to these terms. The stability of the Lorenz model is grounded on an analytical result\cite{Onlong-termboundednessofGalerkinmodels} that assumes certain symmetry within the model, as discussed in the Sec. \ref{ch L05}. Given that our employed estimator does not inherently preserve this symmetry, it gets disrupted in the estimated models even with non-noisy data. In the absence of such symmetry in the estimated model, it becomes challenging to isolate the contributions of model structure change and poor parameter estimation to the emergence of divergent trajectories. Addressing this issue might be within the scope of more advanced SINDy methodologies, such as the Constrained SR3 \cite{ConsSR3} or Trapping SR3 \cite{Promotingglobalstability}, which allow for the incorporation of additional model constraints. However, as these methodologies employ a more sophisticated regression\cite{SR3} than the STLSQ, conducting such an analysis would necessitate a separate study.

\section{CONCLUSION}\label{ch conclusion}

Our work demonstrates a methodology for the estimation of high-dimensional ODE models, assuming an ideal case where there are no hidden variables left unmeasured, and furthermore, a coordinate basis in which the estimated model assumes a sparse representation. Specifically, we considered the task of extracting complex model parameters from time series burdened by noise using the SINDy estimator \cite{SINDy}, with the intention to understand the results, particularly the influence of measurement noise on the dynamical and statistical properties of the corresponding estimated models. To this end, we examined the dependence of dynamical properties of estimated dynamical models, derived using a STLSQ variant of the SINDy algorithm, on the strength of Gaussian noise present in the data. This data were generated by a multidimensional dynamical system, specifically the Lorenz 2005 type II model \cite{L05} (L05 II) in the chaotic regime with a finite attractor.

The dynamical properties of interest were Lyapunov spectrum, Fourier power spectral density and spatial correlation function $C_k^{(2)}$. We found that the dynamical and statistical properties of the estimated models are quantitatively comparable to those of the source system for noise levels at which the two models have a similar attractor in size and space. The Lyapunov spectrum of the estimated models is with increasing noise level moving closer to zero, especially for negative values, resulting in decreasing chaoticity of the estimated model. This is supported by the comparison of power spectral densities obtained in both, source and estimated, models where we can see an increase of power at lower frequency end with increasing noise level. The spatial correlation function shows that as the noise increases, the estimated models increasingly break the translation symmetry inherent in the source system. Nevertheless, the symmetry is broken noticeably just below the value of noise amplitude at which STLSQ algorithm fails to give an estimated model that has a finite global attractor.

Additionally, we studied properties and evaluated the sensitivity of the STLSQ variant of SINDy algorithm by examining how small random perturbations of data affect the model coefficients (through coefficient covariance matrices). We noticed that the estimated model equations of motion for algorithm-intrinsic parameters (threshold $\lambda$ and regularization strength $\alpha$) chosen through the standard cross-validation procedure agree only approximately with the source model equations even for long sampling times, meaning that the STLSQ algorithm has a slight bias.

The covariance matrices of model coefficients turned out to be nearly diagonal for the whole range of tested perturbation amplitudes. They have expected quadratic dependence on noise level in the limit of small perturbations. The corresponding correlation matrices were approximately block diagonal, with the considerable off-diagonal elements corresponding to coefficients belonging to the same or spatially neighboring components of the model. The quadratic trend is broken well below the noise level, where we observe a qualitative change in the set of non-zero estimated model coefficients in comparison to the source model. Correlation matrices in this noise level range show an excessively strong correlation between coefficients belonging to the same estimated ODE components. The reason for such behavior is yet to be understood and will be studied in future work. 

This paper provides a starting point for deeper investigations into the dynamical properties of the estimated dynamical models obtained from noisy data, as well as properties of inference algorithms when applied to such data. While this paper focused on the most physically relevant form of white Gaussian noise, similar analyses could be conducted for other forms of physically relevant noise. Furthermore, similar studies can be carried out for some of the other variants of SINDy, such as SR3 \cite{SR3}, Constrained SR3 \cite{ConsSR3}, and in particular, Trapping SR3\cite{Promotingglobalstability}, which searches for regression solutions in the parameter space that restrict the dynamics of the estimated models to a finite volume of state-space. 

Expanding on this research, it is essential to consider more complex systems where conventional sparse optimization methods might reach their limits. In these scenarios, alternative approaches, such as machine-learning models\cite{kong2023reservoir, wang2019neural}, may provide more effective solutions for system identification and prediction.

\section*{Acknowledgements}
Authors acknowledge the financial support from the Slovenian Research Agency (research core fundings No.~P1-0402, No.~P2-0001 and Ph.D. grant for Aljaž Pavšek).

\section*{AUTHOR DECLARATIONS}

\subsection*{Conflict of interest}

The authors have no conflicts to disclose.

\section*{Data Availability Statement}

The data that support the findings of this study are available from the corresponding author upon reasonable request.

%

\end{document}